\documentclass[oldversion]{aa}
\usepackage{epsfig,psfig}
\usepackage{txfonts}
\usepackage{color}
\usepackage{graphicx}

\newcommand{\msun}{\mbox{$M_{\odot}$}}
\newcommand{\Msun}{\mbox{$M_{\odot}$}}
\newcommand{\lsun}{\mbox{$L_{\odot}$}}
\newcommand{\Lsun}{\mbox{$L_{\odot}$}}

\newcommand{\zsun}{\mbox{$Z_{\odot}$}}
\newcommand{\teff}{\mbox{$T_{\rm eff}$}}

\newcommand{\vinf}{\mbox{$\varv_{\infty}$}}

\newcommand{\mdot}{\mbox{$\dot{M}$}}

\newcommand{\msunyr}{\mbox{$M_{\odot} {\rm yr}^{-1}$}}

\newcommand{\realcleardoublepage}{\clearpage
  \ifodd \arabic{page}\else \thispagestyle{empty}\mbox{}\newpage \fi }

\newcommand{\be}{\begin{equation}}
\newcommand{\ee}{\end{equation}}

\newcommand{\mstar}{\mbox{$M_{\star}$}}

\newcommand{\kmsec}{\mbox{km\,s$^{-1}$}}

\begin{document}

\title{Winds from stripped low-mass helium stars and Wolf-Rayet stars}

\author{Jorick S. Vink\inst{1}}
\offprints{Jorick S. Vink, jsv@arm.ac.uk}

\institute{Armagh Observatory and Planetarium, College Hill, Armagh, BT61 9DG, Northern Ireland}

\titlerunning{Winds from low-mass helium stars and Wolf Rayets}
\authorrunning{Jorick S. Vink}

\abstract{We present mass-loss predictions from Monte Carlo radiative 
transfer models for helium (He) stars as a function of stellar mass, down to 
2\msun. Our study includes both massive Wolf-Rayet (WR) stars and low-mass He stars that 
have lost their envelope through interaction with a companion. For these low-mass He stars 
we predict mass-loss rates that are an order of magnitude smaller than by extrapolation of empirical WR mass-loss 
rates. Our lower mass-loss rates make it harder for 
these elusive stripped stars to be discovered via line emission, and we should attempt to find these stars
through alternative methods instead. 
Moreover, lower mass-loss rates make it 
less likely that low-mass He stars provide stripped-envelope supernovae (SNe) of 
type Ibc. 
We express our mass-loss predictions as a function of $L$ and $Z$ 
and {\it not} as a 
function of the He abundance, as we do not consider this physically astute given our earlier work. 
The exponent of the $\dot{M}$ versus $Z$ dependence 
is found to be 0.61, which is less steep than relationships derived 
from recent empirical atmospheric modelling. Our shallower exponent will make it more challenging to produce ``heavy'' black holes of 
order 40\msun, as recently discovered in the gravitational wave event GW\,150914, making low metallicity  
for these types of events even more necessary.}


\maketitle


\section{Introduction}
\label{s_intro}

We present mass-loss predictions for helium (He) stars, including both massive Wolf-Rayet 
(WR) stars, and low-mass He stars that have presumably lost their hydrogen (H) 
envelope through binary mass transfer (e.g. Podsiadlowski et al. 1992).

One of the main ingredients for massive star evolution modelling concerns their rates of 
stellar wind mass loss (Langer 2012; Yoon et al. 2012; Georgy et al. 2013; 
Limongi \& Chieffi 2006). 
For the H-burning main sequence, we are in the fortunate position that a theoretical recipe 
such as that by Vink et al. (2000, 2001) can be utilized in a meaningful way, but for more 
evolved stars, such as WR stars, the community still needs to rely on empirical recipes, such as 
the recipe by Nugis \& lamers (2000).  
A notable drawback of employing empirical recipes is that they cannot be applied in a parameter space 
for which they have not been derived. These recipes might fail dramatically outside that parameter 
space, for example for low-mass He-stripped stars, especially when the recipe has not been 
constructed from the physically correct stellar parameters (Vink \& de Koter 2002; Puls et al. 2008). 
One of the main aims of this study is to alleviate this shortcoming.  

Reliable mass-loss rates are fundamental to our understanding of the late evolution of massive 
stars towards core collapse, either as supernovae (SNe) Ibc, or direct black hole 
formation (Eldridge \& Vink 2006; Belczynski et al. 2010). 
Such mass-loss rates are especially relevant for understanding merging black holes in recent gravitational wave 
events such as GW\,150914 (Abbott et al. 2016).
These rates are also fundamental for correctly predicting the ionizing radiation of these hot stars with 
major consequences for interpreting He {\sc ii} line emission at intermediate and high red shifts as 
Population {\sc iii} stars (Cassata et al. 2013; Sobral et al. 2015) or very massive stars
(VMS) at low metallicity (Gr\"afener \& Vink 2015; Sz\'ecsi et al. 2015). 

There has been much recent attention to detecting the elusive binary progenitors 
of Ibc supernovae. These objects should be stripped stars that in some sense resemble classical WR stars, but 
because of their weaker winds may either have very little line emission (G\"otberg et al. 2017) or 
no line emission at all. Addressing this question is one of the main aims of this paper. 

Mass loss from massive stars is driven by radiative forces on spectral lines 
(Lucy \& Solomon 1970; Castor, Abbott \& Klein 1975; CAK). 
The CAK work developed the force multiplier formalism  
to treat all relevant ionic transitions. 
This enabled these authors to predict the wind 
mass-loss rate, $\mdot$, and terminal velocity, $\vinf$, simultaneously. 
Using a ``global energy'' Monte Carlo approach (Abbott \& Lucy 1985) in which the velocity 
law was adopted -- aided by empirical constraints -- Vink et al. 
(2000) predicted mass-loss rates for galactic O-type stars that include multi-line scattering.
M\"uller \& Vink (2008) subsequently introduced a new parametrization of the line acceleration, 
expressing it as a function of radius instead of the velocity 
gradient (as in the CAK theory). The implementation of this new formalism improved the
local consistency of Monte Carlo models that initially assumed a velocity law. 
Not only do we find reasonably good agreement with observation of wind terminal velocities, 
but because our method naturally accounts for the physics of multi-line scattering, it is also applicable to 
denser winds, such as those of WR stars (Vink et al. 2011a; Vink 2015).

This paper is organized as follows. 
In Sect.~\ref{s_model}, we briefly describe the Monte Carlo modelling and  
parameter space considered for this study.
The mass-loss predictions (Sect.~\ref{s_res}) are followed by a discussion, 
before ending with a summary in Sect.~\ref{s_sum}.


\section{Monte Carlo models and parameter space}
\label{s_model}

Helium star mass-loss rates are calculated with a 
Monte Carlo method, in which we follow the fate of a large number of 
photon packets ($2~10^6$) from below the stellar photosphere throughout the entire 
wind, up to the terminal velocity.
The core of the method is linked to the total loss of 
radiative energy that is coupled to the momentum gain 
of the outflowing gas (Abbott \& Lucy 1985; Vink et al. 1999).
As the photon absorptions and scatterings 
depend on wind density and hence on the mass-loss
rate $\dot{M}$, we are able to obtain a consistent model where the 
momentum of the gas equals the transferred momentum by photons.
Here we shall employ our local dynamical approach (M\"uller \& Vink 2008, 2014; 
Muijres et al. 2012), in which we predict $\mdot$, $\vinf$, and the wind structure 
parameter, $\beta$, simultaneously.

The Monte Carlo code employs the density and temperature stratifications from 
a model atmosphere calculation performed with {\sc isa-wind} (de Koter et al. 1993). 
These model atmospheres are unified, implying continuity between the 
photosphere and wind region. The models describe the radiative transfer 
in spectral lines adopting an improved Sobolev approximation that includes the diffuse
radiation field. 
The species that are explicitly calculated in non-LTE are 
H, He, C, N, O, S, and Si. The iron-group elements, which are crucial for 
the radiative driving, 
are treated in a generalized version of the modified nebular approximation 
(Schmutz 1991). However, we performed test calculations in which we also 
explicitly computed Fe (iron) in non-LTE. These tests revealed that differences with respect 
to the assumption of the modified nebular approximation for Fe were negligible, and  
we therefore decided to keep the treatment of Fe approximate.           

The line list used for the calculations consists of 
over $10^5$ of the strongest transitions of the elements H~-~Zn 
extracted from the line list constructed by Kurucz \& Bell (1995). 
The wind was divided into 90 concentric shells, with many narrow 
shells in the subsonic region and wider shells in supersonic layers.

Other assumptions involve wind stationarity and sphericity. 
The latter seems to be a reasonable approximation, given that the bulk 
of WR stars are unpolarized (Vink \& Harries 2017). 
However, there are notable exceptions, involving those that are surrounded 
by ejecta nebulae (Vink et al. 2011b). Furthermore, polarization variability due to 
small-scale structure or wind clumping has been 
encountered in luminous blue variables (Davies et al. 2005, 2007).
It has been well established that 
small-scale clumping of the outflowing gas has a pronounced effect 
on the ionization structure of both O-star and WR  
atmospheres (e.g. Hillier 1991). This has lead to a downward revision
of {\em empirical} mass-loss rates, by a factor of $\sim$2-3 (Moffat \& Robert 1994; Puls et al. 2008; 
Hamann et al. 2008; Vink \& Gr\"afener 2012; Ram\'irez-Agudelo et al. 2017).

Moreover, small-scale clumping might have a quantitative effect on the 
radiative driving itself as well, thereby affecting {\em predicted} mass-loss
rates. The subtle issues of clumping and porosity on the mass-loss rate predictions 
were studied by Muijres et al. (2011), who found that whilst 
the impact on \mdot\ can be large for certain specific clumping prescriptions, it was concluded
overall that moderate 
clumping and porosity does not appear to 
change the mass-loss rate predictions substantially.
In the present set of computations we therefore do not
account for the effects of clumping. Nevertheless, it should be kept in mind that clumping might 
affect the predicted values.

\subsection{Parameter range}

We wish to predict mass-loss rates for He stars over a wide mass range, focusing on the 
2-20\,\msun\ range, but also going up to 
60\,\msun\ and down to 0.6\,\msun\ into the regime of the subdwarf O (sdO) stars.
The main focus of this study involves the stripped He stars of order 4\,\msun, as this 
is the mass range of stripped stars that fall just below that of the classical WR stars (in the range 
5 - 20 \msun; Nugis \& Lamers 2000, which should represent absolute \mdot\ values fairly well in the 
WR range).
The luminosities in our study follow from the mass-luminosity relation of Gr\"afener et al. (2011).
This leads to similar values for the low-mass stripped He stars as the models by G\"otberg
et al. (2017).

We do not express $\dot{M}$ as a function of the Eddington parameter, as this would 
depend on the amount of ionized He versus H, 
whilst Vink \& de Koter (2002) and Vink et al. (2011a) showed that $\dot{M}$ should 
hardly depend of the amount of He. However, the effective Eddington factor remains
of key importance for higher mass models in the classical WR range.

The effective temperature sets the ionization stratification in the atmosphere and 
determines which lines are most active in driving the wind. As a result, \teff\
affects the predicted mass-loss rate. Here, we fix \teff\ to 
50\,000\,K, which is characteristic for the best observed example of a stripped He star 
HD\,45166 (Groh et al. 2008). 
We do not express the mass-loss rates of classical WR stars 
as a function of \teff, as radii are highly uncertain because of envelope inflation and 
the possibly associated clumped nature of stellar envelopes (Gr\"afener et al. 2012).

Our grid was constructed to predict the mass-loss behaviour as a function of $L$ 
(or $M$ via the $M$-$L$ relation) and $Z$, which are scaled to the solar values 
(Anders \& Grevesse 1989).

\section{Results}
\label{s_res}

\begin{table}
\centering
\begin{tabular}{lll|ccc}
\hline
\hline\\[-6pt]
\mstar  & $\log L$ & $Z/\zsun$ & \vinf & $\log \mdot$ & \(\beta\) \\[2pt]

[\msun] & [\lsun] &           & [\kmsec] &[\msunyr] &\\[3pt]
\hline\\[-7pt]
0.6    &  1.95  & 1         &    2326 & $-$11.12 &  0.70\\
\hline
2      &  3.49  & 1         &    2650 & $-$8.44  & 0.75\\
       &        & $1/3$     &    2487 & $-$8.90  & 0.78\\
       &        & $1/30$    &     635 & $-$10.18 & 0.54\\
3      &  3.95  & 1         &    2649 & $-$7.82 &  0.79\\
       &        & $1/3$     &    2655 & $-$8.18 &  0.81\\
\hline
4      &  4.26  & 1         &    2871 & $-$7.43 &  0.85\\
\hline
       &        & $1/3$     &    2667 & $-$7.72 &  0.82\\
       &        & $1/10$    &    2252 & $-$8.18 &  0.83\\
5      &  4.48  & 1         &    3039 & $-$7.15 &  0.88\\ 
       &        & $1/3$     &    2787 & $-$7.43 &  0.83\\
       &        & $1/10$    &    2481 & $-$7.80 & 0.89\\      
8      &  4.93  & 1         &    3265 & $-$6.58 & 0.96\\
       &        & $1/3$     &    2714 & $-$6.79 & 0.87\\
       &        & $1/10$    &    2672 & $-$7.11 & 0.88\\
       &        & $1/33$    &    2162 & $-$7.50 & 0.97\\
       &        & $1/100$   &    1489 & $-$8.16 & 1.11\\
10     & 5.13   & 1         &    3472 & $-$6.36 & 1.00\\
       &        & $1/3$     &    2706 & $-$6.55 & 0.87\\  
       &        & $1/10$    &    2655 & $-$6.85 & 0.87\\
       &        & $1/33$    &    2319 & $-$7.29 & 0.96\\
15     & 5.47   & 1         &    3790 & $-$5.94 & 1.04\\
       &        & $1/3$     &    2848 & $-$6.08 & 0.92\\
       &        & $1/10$    &    2587 & $-$6.33 & 0.88\\
       &        & $1/33$    &    2485 & $-$6.69 & 0.93\\
       &        & $1/100$   &    1985 & $-$7.13 & 1.08\\
20     & 5.69   & 1         &    3960 & $-$5.67 & 1.08\\ 
       &        & $1/3$     &    2953 & $-$5.80 & 0.96\\
       &        & $1/10$    &    2576 & $-$6.04 & 0.90\\
       &        & $1/33$    &    2504 & $-$6.35 & 0.92\\
       &        & $1/100$   &    2111 & $-$6.78 & 1.03\\
       &        & $1/333$   &    1739 & $-$7.418& 1.17\\
\hline
60     & 6.40   & 1         &    4616 & $-$4.80 & 1.28\\ 
       &        & $1/3$     &    3507 & $-$4.94 & 1.10\\
       &        & $1/10$    &    2629 & $-$5.09 & 0.95\\
       &        & $1/33$    &    2347 & $-$5.31 & 0.91\\
       &        & $1/100$   &    2308 & $-$5.63 & 0.93\\
\hline
\end{tabular}
\caption{Mass-loss predictions for the He main sequence. 
\teff\ is kept constant at 50,000.}
\label{tab:results}
\end{table}

The main aim of the paper is to get a handle on the mass-loss rates for stripped 
He stars with masses just below those of classical WR stars, i.e. at about 4\msun.
We predict a rate of $\log \mdot = -7.43$ and a terminal wind velocity of 2871 km/s for our 4\msun\ model. 
This mass-loss rate is an order of magnitude {\it smaller} than that assumed 
by G\"otberg et al. (2017) through extrapolation of empirical WR stars by Nugis \& Lamers (2000).

\begin{figure}
\centerline{\psfig{file=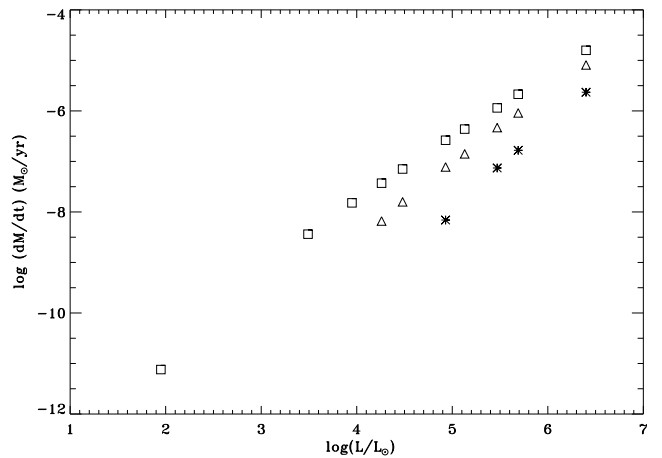, width = 9 cm}}
\caption{Predicted mass-loss rates vs. $\log(L/\Lsun)$ for solar metallicity models (open squares), 
models of 10\% \zsun\ (open triangles), and 1\% \zsun\ (asterisks). The high-mass models of 60\Msun\ follow 
the bulk of the models very well; the very low-mass models with M=0.6\Msun\ ($\log(L/\Lsun) = 1.95$) fall somewhat
below the general behaviour.}
\label{f_mdot}
\end{figure}

Table~\ref{tab:results} lists the rest of our mass-loss predictions. 
The initial stellar parameter columns are self-explanatory.
The resulting predicted wind terminal velocities, mass-loss rates, and 
wind acceleration parameter $\beta$ are 
given in columns (4), (5), and (6), respectively. 
The predicted mass-loss rates (column 5) are shown in Fig.~\ref{f_mdot}. 
Various symbols are used to identify various $Z$ ranges.
For visual clarity, not all $Z$ models feature here, but only a subset -- with alternating 
$Z$ values -- from Table~1 were plotted.

Figure 1 shows that $\mdot$ increases with $L$ and $Z$, as expected. Whilst the very high-mass model (with
$M$ = 60\msun; $\log(L/\lsun) = 6.4$) fits the relationship formed by the bulk of models in the 2-20\msun\ range 
($\log(L/\lsun) = 3.49 - 5.69$ range) very well, the very 
low-mass model (M=0.6\Msun; $\log(L/\Lsun) = 1.95$) with  $\log \mdot = -11.12$ falls 
below the general behaviour. It will therefore not be included in the mass loss versus $L$ power-law fitting.

When we compare the $M = 0.6\,\msun$ model to the predictions of Vink \& Cassisi (2002) for horizontal
branch stars, we find $\log \mdot = -10.40$. However, the new dynamically consistent rate has \vinf\ 
a factor 2 higher, and for a constant wind momentum one would expect an \mdot\ value a factor $\sim 2$ 
lower, i.e. $\log \mdot \simeq -10.70$. This is within a factor of two from the current value.
Our value for the 0.6\msun\ model is a factor of $\sim$5 larger than the SdO mass-loss predictions by 
Krti{\v c}ka et al. (2016).

We next turn our attention to the wind velocity structure parameter, $\beta$, which 
describes how rapidly the wind accelerates. 
The predicted values of $\beta$ are shown in column (6) of the Table. 
$\beta$ shows a slight dependence on stellar mass, but mostly $\beta$ values are of order unity, in accordance 
with the O-star models of Pauldrach et al. (1986), M\"uller \& Vink (2008), and Muijres et al. (2012). 

\subsection{Mass-loss recipe for He stars}
\label{sec_gammae}

In order to determine the dependence of the mass-loss rate on $L$ and $Z$ simultaneously,
we perform multiple linear regression, and we find
 
\begin{equation}
\label{eq:fit}
\log \mdot~=~-13.3~+~1.36 \log(L/\lsun)~+~0.61 \log(Z/\zsun)    
\end{equation}
with a fitting error of $\sigma$ = 0.11. The formula was derived for the $Z$ 
range $(Z/\zsun) = 1 - 10^{-2}$ and the mass range 2 - 60\,\msun, but for reasons given earlier 
application of this formula is recommended only for lower-mass stripped stars, and not for
classical WR stars.

The mass-loss versus mass relationship (as explored by e.g. Langer 1989) 
can be transformed using relevant mass-luminosity relationships (e.g. Gr\"afener et al. 2011).

\section{Discussion}
\label{disc}

Comparing our mass-loss predictions against observed mass-loss rates 
is a non-trivial undertaking, as low-mass stripped He stars are as yet elusive. 

Arguably the best example of a stripped star is HD\,45166, which
consists of a 4\msun\ quasi-WR (qWR) star orbiting a B7V companion (Groh et al. 2008), which
may have evolved from a 12\msun\ primary (G\"otberg et al. 2017). HD\,45166 has a surprisingly low
(equatorial)) terminal wind velocity of 350\,km/s. For such a low \vinf, and employing our 
global approach (Vink et al. 2000), we would arrive at a mass-loss rate that is an order of magnitude {\it larger} than when we use 
a higher value, such as for the presumed polar wind component. 
The agreement between the observed mass-loss rate from HD\,45166 from Groh et al. (2008) and the 
mass-loss rates given by extrapolation of the empirical Nugis \& Lamers (2000) WR recipe
is probably the main reason why G\"otberg et al. (2017) assumed that the 
Nugis \& Lamers (2000) WR mass-loss rates provide rates in the expected range for low-mass He 
stars, whilst we find that they are actually overestimated by an order of magnitude.

Regarding the metallicity dependence, the exponent of 0.61 is slightly less than 
the predicted value of 0.85 in Vink \& de Koter (2005). This is to be expected, as the
earlier predictions were based on the global approach, and the difference may simply be attributed 
to the dependence of \vinf\ on $Z$ (Vink et al. 2001). 

Our $\dot{M}$-$Z$ exponent is also smaller than that derived in the recent empirical study by Hainich et al. (2015), who suggest
 an $\dot{M}$-$Z$ exponent larger than unity. 
Given that, firstly, the Hainich et al. empirical rates were 
derived for a data set showing a significant scatter 
when expressed as a function of $L$, which suggests the Hainich et al. (2015) study misses 
a relevant parameter and, secondly, that our exponent was derived 
over a far wider range in $Z$ than the Hainich et al. data, 
we consider our lower exponent of 0.61 to be more likely close-to-correct.
  
Note that these $\dot{M}$-$Z$ exponents are predominately determined by the Fe abundance, and 
distinct from the $Z$-dependence in the empirical Nugis \& Lamers (2000) recipe. We do not 
consider the Nugis \& Lamers (2000) formulation physically astute, but it should still 
provide reasonably accurate \mdot\ estimates within the actual classical WR range.

\section{Summary}
\label{s_sum}

We presented mass-loss predictions from Monte Carlo radiative 
transfer models for stripped He stars by stellar winds or mass transfer through a companion.

\begin{itemize}

\item{The mass-loss rate is expressed as a function of $L$ and $Z$ through 
multiple linear regression}.
 
\item{We find $\log$ \mdot\ $=$ $-$13.3 $+$ 1.36 $\log(L/\lsun)$ $+$ 0.61 $\log(Z/\zsun)$, 
derived for $(Z/\zsun)$ $=$ $1 - 10^{-2}$.}

\item{Our mass-loss rates for low-mass stripped stars are an order of magnitude lower
than would be found by extrapolating empirical WR recipes.}

\end{itemize}

This has major implications for the search for low-mass stripped He stars as the progenitors of SNe Ibc and ionizing sources
in the Universe.





\end{document}